\documentclass[prc,twocolumn,showpacs,amsmath,amssymb]{revtex4}

\usepackage{graphicx}
\usepackage{dcolumn}
\usepackage{bm}

\begin{document}

\title{Modelling of compound nucleus formation in fusion of heavy nuclei}

\author{A. Diaz-Torres}
\affiliation{Institut f\"ur Theoretische Physik der
Johann Wolfgang Goethe--Universit\"at Frankfurt, Robert Mayer 10,
D--60054 Frankfurt am Main, Germany and \\
Institut f\"ur Theoretische Physik der
Justus--Liebig--Universit\"at Giessen, Heinrich--Buff--Ring 16,
D--35392 Giessen, Germany}

\date{\today}

\begin{abstract}
A new model that includes the time-dependent dynamics of the single-particle (s.p.)
motion in conjunction with the macroscopic evolution of the system is proposed for describing
the compound nucleus (CN) formation in fusion of heavy nuclei.
The diabaticity initially keeps the entrance system around its contact configuration, but the
gradual transition from the diabatic to the adiabatic potential energy surface (PES) leads to fusion or
quasifission. Direct measurements of the probability for CN formation are crucial to
discriminate between the current models.

\end{abstract}



\maketitle


The understanding of new experimental results on fusion of heavy nuclei and the formation of
superheavy elements (SHE) require modelling, not only of the initial capture process and the final
CN de-excitation process, but also of the intermediate stage of evolution of the combined system
from the contact configuration into the CN. The competition between fusion and quasifission
(reseparation before CN formation) can inhibit fusion by many orders
of magnitude, e.g. \cite{Reisdorf}. Understanding this inhibition may be the key to forming more
SHE. Nowadays, there is no consensus \cite{Volkov} for the mechanism of the CN formation
in fusion of heavy nuclei near the Coulomb barrier.
Depending on the main coordinate for fusion, two sorts of models can be distinguished.
In the first type \cite{Greiner,Swiatecki,Aguiar,Aritomo,Tokuda,Denisov,Zagrebaev1,Shen,Abe1}, the
fusion occurs along the radial coordinate using adiabatic PES obtained either with the
liquid drop model or Strutinsky's macroscopic-microscopic method.
The competition between fusion and quasifission, which depends on the fluctuactions \cite{Abe2},
has only been included in recent models, e.g. \cite{Zagrebaev1,Shen,Abe1}.
However, the experimental data are not always explained.
In the second type \cite{DNS2} (dinuclear system (DNS) model), the fusion happens in the
mass asymmetry coordinate
$\eta=(A_1-A_2)/(A_1+A_2)$  where $A_1$ and $A_2$ are the mass numbers of the nuclei.
The DNS nuclei remain at the touching configuration and exchange nucleons until either
all nucleons have been transferred
from the lighter to the heavier fragment (complete fusion), or the DNS decays
before the CN formation (quasifission). The model assumes a sudden (double-folding in frozen
density approximation) PES in the radial
coordinate, while the PES behaves adiabatically along the fusion path in the $\eta$
coordinate. Although the model has been used to explain many experimental evaporation
residues (ER) cross sections, its theoretical foundation is not clear enough yet.

In this paper a new model is proposed for the CN formation, which is based
on the following
general ideas (which are well-established but have up to now not been used in combination
in any of the current models of fusion):
(i) Once the two nuclei are at the
contact point, the system moves in a multidimensional space of collective coordinates
$\textbf{q}$ (shape parameters),
(ii) this motion is governed by the master equation, and (iii) the nature of the
s.p. motion is time-dependent, it is initially diabatic and then
approaches the adiabatic limit due to residual two-body collisions.
Consequently, the system moves in a time-dependent PES, $V(\textbf{q},t)$, which is initially
diabatic and gradually becomes adiabatic. In the diabatic limit \cite{Nore}
(elastic nuclear matter), the nucleons
do not occupy the lowest free s.p. energy levels as in the adiabatic case
(plastic nuclear matter), but keep
their quantum numbers and remain in the diabatic states during a collective motion of the system.
This approach is realistic in the initial stage of collisions near
the Coulomb barrier where the total excitation energy per nucleon $E^{*}\gtrsim 0.03$ MeV
\cite{Cass,Alexis1}.
During the transition from the diabatic to the adiabatic limit, the nuclear matter is
elastoplastic like glycerine.
In contrast to the current models for CN formation, the present approach explicitly includes
for the first time the time-dependent dynamics of the s.p. motion in conjunction with the
macroscopic evolution of the system into the CN.
In Refs. \cite{Aritomo,Tokuda,Zagrebaev1,Shen}
the ideas (i) and (ii) were applied, but the authors exclusively used adiabatic PES.
In a recent paper \cite{Alexis1}, only idea (iii) was used to calculate the dynamical
potential for the radial motion of the combined system, while in the past only the diabatic
limit of the s.p. motion was used to study the initial capture process
\cite{Cassing2,Berdichevsky}.
The ideas (i)-(iii) have solely been applied in combination to describe
deep-inelastic reactions \cite{Yadav}.
The scenario presented here for the CN formation shows that following contact
the diabaticity forms a valley in the PES where the system
remains trapped around its touching configuration, but the gradual transition from the diabatic to
the adiabatic PES allows the system to evolve in shapes leading to fusion or quasifission.
The timescale for the decay to the adiabatic PES is crucial in determining the timescale of the
transition to the compact fused system. The calculations are based on the master equation
\cite{Norenberg} and the diabatic two-center shell model (TCSM) \cite{Lukasiak2} developed using
the asymmetric TCSM (ATCSM) \cite{ATCSM}. The critical ingredients of the model are (i) temperature,
(ii) diabatic PES, (iii) adiabatic PES, (iv) transition from diabatic to adiabatic PES,
and (v) shape parameters. The evaluation of each will be described, then the results from the model
will be presented.

The time-dependent population probabilities $p$ for the different configurations $\textbf{q}$
(shapes) of the system are solutions of the master equation

\begin{equation}
\dot p(\textbf{q},t)=\sum_{\textbf{q}^{'}}[\Lambda(\textbf{q},\textbf{q}^{'},t)
p(\textbf{q}^{'},t)-\Lambda(\textbf{q}^{'},\textbf{q},t)p(\textbf{q},t)], \label{eq_6}
\end{equation}
where the macroscopic transition probabilities according to
Ref. \cite{Moretto} are
$\Lambda(\textbf{q},\textbf{q}^{'},t)=\kappa_{0}exp[V(\textbf{q}^{'},t)/2T(\textbf{q}^{'},t)-
V(\textbf{q},t)/2T(\textbf{q},t)]$,
justified by the assumption that the level density of the system
determines the transition. The strengh constant $\kappa_0$ characterizes
the global time scale and has a realistic value of $\sim10^{22}$ s$^{-1}$ \cite{Gippner}.
The sum in Eq. (\ref{eq_6}) is extended only to the nearest configurations $\textbf{q}^{'}$
(the collective coordinate space is discretized).
It is assumed that the system is initially at the contact configuration $\textbf{q}_0$
where the s.p. occupation numbers obey a Fermi-distribution $n_{\alpha}^{F}(\textbf{q}_0,T_0)$
for a temperature $T_0$.
Configurations other than $\textbf{q}_0$ are not populated at this time, and hence
the initial condition for Eq. (\ref{eq_6}) is $p(\textbf{q},0)=\delta_{\textbf{q},\textbf{q}_0}$.
The temperature $T_0$ is related
to the excitation of the system immediately after the capture process. This temperature
can be estimated either as $T_0\approx\sqrt{[E_{c.m.}-V(\textbf{q}_0,0)]/a}$, where $E_{c.m.}$ is the
total incident energy in the center of mass frame and $a=A/12$ MeV$^{-1}$ ($A$ is the total mass number of the system),
if the initial radial kinetic energy is dissipated when
the nuclei reach the contact point, or using a frictional model, e.g. \cite{Shen}, for the capture process.
The local (at fixed $\textbf{q}$) temperature $T=\sqrt{E_{exc}/a}$ is defined by means of
the local excitation energy

\begin{equation}
E_{exc}(\textbf{q},t)=a\,T_0^2 +
\int^{t}_{0}{(- \frac{d \Delta V_{diab}(\textbf{q},t^{'})}{d t^{'}})
p(\textbf{q},t^{'}) dt^{'}}, \label{eq_3}
\end{equation}
which results from the decay of the diabatic part $\Delta V_{diab}$ of the potential $V$.
$\Delta V_{diab}$ represents an energetic hindrance for the initial system to reach a
configuration $\textbf{q}$, if the nucleons follow diabatic levels during this process
(elastic response).
The local excitation of the system is caused by the loss of its elasticity \cite{Nore}.
$\Delta V_{diab}$ is calculated as

\begin{equation}
\Delta V_{diab}(\textbf{q},t)\approx\sum_{\alpha}\varepsilon_{\alpha}^{diab}(\textbf{q})
[n_{\alpha}(\textbf{q},t) - n_{\alpha}^{F}(\textbf{q},T)] , \label{eq_4}
\end{equation}
where $\varepsilon_{\alpha}^{diab}$ are the diabatic levels with occupations $n_{\alpha}(\textbf{q},t)$
and $\alpha$ denotes the quantum numbers of these states. The diabatic levels
$\varepsilon_{\alpha}^{diab}$ and their
wave functions $\phi_{\alpha}^{diab}(\textbf{r})$ are obtained with the maximum symmetry
method \cite{Lukasiak2}.
The dynamical PES is defined as
$V(\textbf{q},t)=V_{adiab}(\textbf{q},T)+\Delta V_{diab}(\textbf{q},t)$, where
$V_{adiab}$ is the adiabatic PES which is calculated using Strutinsky's
macroscopic-microscopic method \cite{ATCSM}.
The nuclear part of the macroscopic energy is
obtained with the Yukawa-plus-exponential method \cite{Krappe}. The diabaticity destroys
the Fermi-distribution of the occupations, but the residual two-body collisions gradually
recover it. This process is locally described by the relaxation equation \cite{Cassing}

\begin{equation}
\dot n_{\alpha}(\textbf{q},t)=-\tau^{-1}(\textbf{q},t)[n_{\alpha}(\textbf{q},t) -
n_{\alpha}^{F}(\textbf{q},T)] , \label{eq_1}
\end{equation}
where $\tau$ is an average relaxation time (in order to conserve the number of particles).
The initial occupations $n_{\alpha}(\textbf{q},0)$ are the diabatic ones obtained from
$n_{\alpha}^{F}(\textbf{q}_0,T_0)$. $\tau$ is defined as

\begin{equation}
\tau^{-1}(\textbf{q},t)=\frac{ \sum_{\alpha}[n_{\alpha}(\textbf{q},t) -
n_{\alpha}^{F}(\textbf{q},T)]\,\Gamma_{\alpha}(\textbf{q},T) }
{ N_{coll}\,\hbar\,\sum_{\alpha}n_{\alpha}(\textbf{q},t) } , \label{eq_2}
\end{equation}
where $\Gamma_{\alpha}$ are the widths of the s.p. levels.
The factor $N_{coll}$ is the average number of two-body collisions
per nucleon to establish the equilibrium occupations $n_{\alpha}^{F}$.
The value $N_{coll}=3$ \cite{Cass} will be used.
The expression (\ref{eq_2}) follows the idea that the relaxation process becomes slower when
the occupations approach $n_{\alpha}^{F}$. If the equilibrium was reached,
the relaxation time would be infinite, i.e. the occupations would remain the same.
The widths $\Gamma_{\alpha}$ are obtained with the parametrization
given in Ref. \cite{Helmut}. Since the diabatic s.p. excitations occur around the Fermi surface,
the values $\Gamma_0^{-1}=0.061$ MeV$^{-1}$ for half saturation density and $c=20$ MeV will be used
\cite{Helmut}.

The collective coordinates $\textbf{q}$ are the shape parameters of
the ATCSM \cite{ATCSM}: (i) the elongation $\lambda=l/2R_0$, which measures the length $l$ of the
system in units of the
diameter $2R_0$ of the spherical CN  and describes the relative motion,
(ii) the deformation $\beta_i=a_i/b_i$ of the fragments, defined by the ratio of
their semiaxes,
(iii) the neck coordinate $\epsilon=E_0/E^{'}$, defined by the ratio of the
actual barrier height $E_0$ to the barrier height $E^{'}$ of the two-center oscillator,
and (iv) the volume asymmetry of the nuclear shapes
(equipotential shapes) $\xi=(V_1-V_2)/(V_1+V_2)$, where $V_1$ and $V_2$ are the
volumes of the left and right regions divided by a plane at the necks between the fragments.
The collective coordinate space is divided into three regions: (i) compact shapes around the
spherical shape (fusion region),
(ii) elongated shapes outside the initial contact configuration and beyond the
Coulomb barrier (quasifission region), and (iii) intermediate shapes which could lead to fusion or
quasifission. In the fusion region the physical mass asymmetry $\eta$
(defined like $\xi$ but in terms of the masses which are calculated using
the microscopic density distribution
 $\rho(\textbf{r})=\sum_{\alpha}n_{\alpha}|\phi_{\alpha}^{diab}(\textbf{r})|^{2}$ )
reaches a minimal plateau,
i.e. a maximal number of nucleons move (their wave-functions spread) in the whole volume of the system due to
the decrease of the barrier $E_0$ between the fragments.
The fusion ($P_{CN}$) and quasifission ($P_{QF}$) probabilities are defined as the
sum of the population probabilities $p$ in the fusion and quasifission regions, respectively.
In addition the CN excitation energy $E_{CN}$ is defined as the average
of the excitation energies (\ref{eq_3}) of the different shapes in the fusion region.

In the following the fragments are
considered as spherical ($\beta_i$=1) and the neck coordinate is fixed at $\epsilon=0.75$.
With this value of
$\epsilon$ the Coulomb barrier of the diabatic potential for an initial system with $T_0=0$ MeV
is close to the barrier of the double-folding potential \cite{Alexis1}.
For computational reasons the calculations are done using the
coordinates $\lambda$ and $\xi$ that are the relevant ones in the current models for
CN formation. The non-linear set of equations (\ref{eq_6}-\ref{eq_2}) are solved by
sucessive iterations using a small time step $\Delta t = 10^{-23}$ s.
The master equation (\ref{eq_6}) is solved on a grid
($1\leqslant\lambda\leqslant1.8$,$|\xi|\leqslant0.7$ where $\Delta\lambda = 0.02$ and
$\Delta\xi = 0.1$) with appropriate boundary conditions to follow the continuous split of the
initial population probabilities $p(\textbf{q},0)$ into fusion and quasifission.
Values of $|\xi|>0.7$
are not included in the calculation because the TCSM used \cite{ATCSM} is not appropriate for
large $\xi$. The fusion and quasifission processes are determined at the timescale $t_0$
when $P_{CN}(t_0) + P_{QF}(t_0)\simeq 1$. The model will be applied to some (near)
symmetric central ($l=0$) collisions. In these calculations the excitation energy of the initial
system at the contact configuration is 40 MeV.

\begin{figure}
\includegraphics[width=7.8cm]{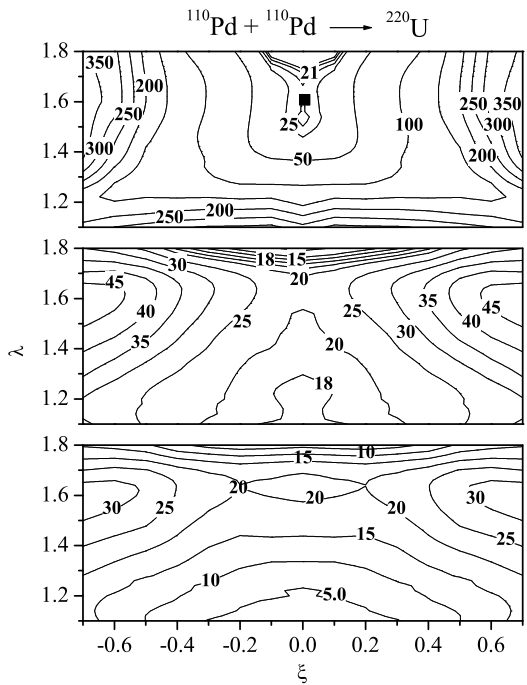}
\caption{Dynamical PES for $^{110}$Pd+$^{110}$Pd: (upper part) at t=0 s (diabatic PES),
the square denotes the contact configuration; (middle part) at $t_0= 4\times10^{-20}$ s;
and (lower part) the adiabatic PES. See text for further details.}
\end{figure}

\begin{figure}
\includegraphics[width=7.8cm]{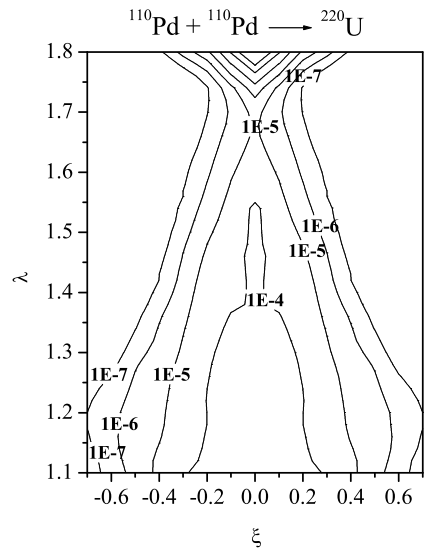}
\caption{Distribution of the population probabilities $p(\lambda,\xi,t_0)$ for
$^{110}$Pd+$^{110}$Pd on the PES of Fig. 1 (middle part).
The fusion region is $\lambda\leqslant1.3$. See text for further details.}
\end{figure}

\begin{table}
\caption{$P_{CN}$, $t_0$ ($10^{-20}$ s) and $E_{CN}$ (MeV) for some (near) symmetric central collisions.
The $P_{CN}$ values are compared to direct experimental $P_{CN}^{\exp.}$ \cite{Hinde}.}
\begin{ruledtabular}
\begin{tabular}{ccccc}
Reaction   & $P_{CN}$   & $t_0$   & $E_{CN}$   & $P_{CN}^{\exp.}$\\
\hline
$^{90}$Zr+$^{90}$Zr$\rightarrow$$^{180}$Hg    & 3$\times 10^{-1}$     & 5    & 40   & \\
$^{100}$Mo+$^{100}$Mo$\rightarrow$$^{200}$Po  & 1.4$\times 10^{-1}$   & 4.5  & 40   & \\
$^{110}$Pd+$^{110}$Pd$\rightarrow$$^{220}$U   & 1.7$\times 10^{-2}$   & 4    & 40   & \\
$^{100}$Mo+$^{110}$Pd$\rightarrow$$^{210}$Ra  & 7.5$\times 10^{-2}$   & 4.3  & 40   & \\
$^{96}$Zr+$^{124}$Sn$\rightarrow$$^{220}$Th   & 4.5$\times 10^{-2}$   & 4    & 40   & 5$\times 10^{-2}$\\
\end{tabular}
\end{ruledtabular}
\end{table}

Fig. 1 shows the dynamical PES for $^{110}$Pd+$^{110}$Pd (i) at
the initial moment (diabatic PES, upper part) when the nuclei are at the contact configuration (square),
(ii) at $t_0= 4\times10^{-20}$ s (middle part), and (iii) the adiabatic PES (lower part).
The PES is normalized with the macroscopic energy of the spherical CN.
In Fig. 2, the distribution of the population probabilities $p(\lambda,\xi,t_0)$ on the PES of Fig. 1
(middle part) is presented. From these figures, it is observed that
the diabaticity initially forms a valley confining the entrance system around its touching
configuration as in the DNS model \cite{DNS2}.
Nevertheless, the gradual transition to the adiabatic PES leads to fusion mainly in the
elongation $\lambda$ (relative distance) or the population of neighbouring configurations
ending in the quasifission channel. In contrast to the DNS model,
the diffusion along the $\xi$ coordinate, while the nuclei remain at the contact point
($\lambda=1.5-1.6$), is strongly suppressed by a large diabatic hindrance.
$P_{CN}$ and $P_{QF}$ are determined before the dynamical PES reaches the adiabatic one
(comparing middle and lower parts in Fig. 1).
The competition between fusion and quasifission is regulated by
the dynamical PES. The probability of configurations which stay close to that of
the entrance channel and reseparate is very large, as can be seen from Fig. 2.
This causes the fusion hindrance for $^{110}$Pd+$^{110}$Pd which is consistent with
the experimental conclusions drawn in Ref. \cite{Reisdorf}.
The final population probabilities
in the fusion region ($\lambda\leqslant1.3$ in Fig. 2) are very small,
and as a result of the expression (\ref{eq_3}), $E_{CN}$ is practically the same as
the excitation energy of the initial system at the contact point, i.e. 40 MeV.
The scenario seen so far is the same for
other systems studied. Table I shows $P_{CN}$, $t_0$ and $E_{CN}$ calculated for some (near)
symmetric reactions. The values of the model
parameters are the same for all the reactions.
The decrease of the $P_{CN}$
values from $^{90}$Zr+$^{90}$Zr to $^{110}$Pd+$^{110}$Pd is observed along with an increase
of the quasifission probability beyond $90\%$. This is because (i) the repulsive character of the diabatic
PES increases with increasing mass number $A$ of the combined system and (ii) the valley of the diabatic
PES becomes more shallow with increasing $A$. Consequently the timescale $t_0$ of
the fusion and quasifission processes decreases with increasing $A$.
The $P_{CN}$ for $^{96}$Zr+$^{124}$Sn agrees very well with the recent direct experimental measurement
\cite{Hinde}. The $P_{CN}^{\exp.}$ includes the contribution of higher partial waves $l$, i.e. $l>0$, although
the main contribution results from low $l$ for heavy systems at this energy. The $P_{CN}$ obtained with
the DNS model for this reaction \cite{DNS2} also agrees well with that experiment, while $P_{CN}$ for, e.g.
$^{110}$Pd+$^{110}$Pd, is about one order of magnitude smaller than the present $P_{CN}$ (see Table I).
The fluctuation-dissipation model in Ref. \cite{Tokuda} predicts $P_{CN}$ values that are much larger than the
present ones, i.e. about one order of magnitude larger for $^{110}$Pd+$^{110}$Pd and $^{100}$Mo+$^{110}$Pd.
Direct experimental measurements of $P_{CN}$ like in Ref. \cite{Hinde} are crucial to
discriminate between the current models for CN formation. The dependence of $P_{CN}$
on the parameter $\Gamma_0^{-1}$ is strong for the heaviest systems studied,
i.e. changing the latter from 0.061 MeV$^{-1}$ to 0.03 MeV$^{-1}$,
causes the $P_{CN}$ decrease by about one order of magnitude due to the faster transition to the
adiabatic PES, whereas the effect on the timescale $t_0$ is rather weak. The timescale $t_0$ is mainly
determined by the quasifission process that occurs near the contact configuration of
the initial system, where the diabatic effects are small.
The dependence of $P_{CN}$ on the excitation energy of the initial system
at the touching point is rather weak, i.e. a saturation of the $P_{CN}$ practically occurs
from $\sim 20$ MeV upwards.

In summary, a new realistic model for the CN formation in fusion of massive nuclei has been
developed. It incorporates for the first time important physical effects which critically affect
the evolution of the fusing system.
The diabaticity initially keeps the entrance system around its touching point,
but the gradual transition from the diabatic to adiabatic PES leads to fusion (mainly in the relative
distance) or quasifission. The dynamical PES regulates the competition between fusion and quasifission.
The probabilities for CN formation in some (near) symmetric central collisions have been obtained and
found to agree very well with the recent direct experimental determination for $^{96}$Zr+$^{124}$Sn.
Direct measurements of the $P_{CN}$ like in Ref. \cite{Hinde} along  with
distributions of the quasifission fragments are crucial to discriminate between the
current models for CN formation.
To calculate ER cross sections the present approach
should be combined with other models that describe the initial capture stage and the survival of the
CN against fission.
A TCSM more appropriate for reactions with large mass-asymmetry is being developed which should
be useful in predicting production cross sections of SHE.


The author thanks W. Scheid, W. Cassing, G.G. Adamian and
N.V. Antonenko for fruitful discussions, M. Dasgupta and C. Beck for a careful
reading of the manuscript and useful suggestions, and
the Alexander von Humboldt Foundation for financial support.

\end{document}